\documentclass[12pt]{iopart}

\usepackage{iopams,graphicx,amssymb}
\eqnobysec

\begin{document}

\title[Random interface growth in random environment]
{Random interface growth in random environment:
Renormalization group analysis of a simple model}

\author{N. V. Antonov and P. I. Kakin}

\address{Department of Theoretical Physics, St.~Petersburg University,
Uljanovskaja 1, St.~Petersburg, Petrodvorez, 198504 Russia}

\ead{n.antonov@spbu.ru, p.kakin@spbu.ru}

\begin{abstract}
We study effects of turbulent mixing on the random growth of an interface
in the problem of the deposition of a substance on a substrate. The growth
is modelled by the well-known Kardar--Parisi--Zhang model. The turbulent
advecting velocity field is modelled by the Kraichnan's rapid-change
ensemble: Gaussian statistics with the correlation function
$\langle vv\rangle \propto \delta(t-t') \, k^{-d-\xi}$, where $k$ is the
wave number and $0<\xi<2$ is a free parameter. Effects of compressibility
of the fluid are studied. Using the field theoretic renormalization group we
show that, depending on the relation between the exponent $\xi$ and the
spatial dimension $d$, the system reveals different types of large-scale,
long-time asymptotic behaviour, associated with four possible fixed points
of the renormalization group equations. In addition to known regimes
(ordinary diffusion, ordinary growth process, and passively advected scalar
field), existence of a new nonequilibrium universality class is established.
Practical calculations of the fixed point coordinates, their regions of
stability and critical dimensions are calculated to the first order of the
double expansion in $\xi$ and $\varepsilon=2-d$ (one-loop approximation).
It turns out that for incompressible fluid, the most realistic values
$\xi=4/3$ or 2 and $d=1$ or 2 correspond to the case of passive scalar field,
when the nonlinearity of the KPZ model is irrelevant and the
interface growth is completely
determined by the turbulent transfer. If the compressibility becomes strong
enough, the crossover in the critical behaviour occurs, and these values of
$d$ and $\xi$ fall into the region of stability of the new regime, where the
advection and the nonlinearity are both important. However, for this regime
the coordinates of the fixed point lie in the unphysical region, so its
physical interpretation remains an open problem.
\end{abstract}


\section{Introduction and description of the model} \label{sec:Intro}

Over decades, constant interest has been attracted to the growth processes
in various physical systems: solidification and flame fronts, smoke and
colloid aggregates, tumors, and so on; see e.g. \cite{rost1}--\cite{ball} and
references therein. A most prominent example is provided by the deposition
of a substance on a substrate and growth of the corresponding phase boundary
(interface). A number of microscopic models were proposed to describe those
phenomena: Eden model \cite{Eden}, Edwards--Wilkinson model \cite{EW},
restricted solid-on-solid model \cite{SOS}, ballistic deposition \cite{ball};
not an exhaustive list.

It turns out, however, that the growth processes share some important
features with those of equilibrium nearly-critical systems: namely,
self-similar (scaling) behaviour (power-like dependencies) with rather
universal (independent of the details of a specific process) exponents.
In particular, the $n$-th order structure functions of a growth process
behave as \cite{rost1}--\cite{rost3}
\begin{equation}
S_{n}(t,r) \equiv
\langle\left[h(t,{\bf x}) - h(0,{\bf 0})\right]^{n}\rangle \simeq
r^{n\chi}\, F_{n} (tr^{z}), \quad r=|{\bf x}|.
\label{scaling}
\end{equation}
Here $h(x)=h(t,{\bf x})$ is the height of the interface profile, the
brackets $\langle\dots\rangle$ denote averaging over the statistical
ensemble, $\chi$ and $z$ are referred to as the roughness exponent and
the dynamical exponent, respectively, and $F_{n}(\cdot)$ is a certain
universal scaling function. The asymptotic behaviour (\ref{scaling})
takes place in the infrared (IR) range, where the time and space
differences $t$, $r$ are large in comparison with characteristic
microscopic scales.

It is then natural to try to describe universal properties of the growth
processes on the base of a certain simplified model for a smoothed
(coarse-grained) height field, in analogy with the theory of critical state,
where most typical universality classes (types of critical behaviour)
are described by the classical $\varphi^{4}$-model \cite{Zinn,Book3}.
As the coarse-grained model of growth, one usually choses the
Kardar--Parisi--Zhang (KPZ) model \cite{KPZ}, described by the nonlinear
stochastic differential equation
\begin{equation}
\partial_{t} h= \varkappa_0\, \partial^{2} h +
\lambda_{0}(\partial h)^{2}/2 + f.
\label{KPZ}
\end{equation}
Here the height field $h(x)=h(t,{\bf x})$ depends on the
$d$-dimensional substrate coordinate ${\bf x}$,
$\partial_{t}= \partial/\partial t$,
$\partial_{i}= \partial/\partial x_{i}$,
$\partial^{2}=\partial_{i}\partial_{i}$ is the Laplace operator and
$(\partial h)^{2}=\partial_{i}h\partial_{i}h$; the summations over
repeated tensor indices are always implied. The first term in the
right-hand side of (\ref{KPZ}) describes the surface tension with the
coefficient $\varkappa_0>0$. The second term represent an excess growth
along the local normal to the surface. The parameter $\lambda_{0}$ can
be of either sign; it can be scaled out, and in the following we set
$\lambda_{0}=1$.

Furthermore, $f=f(x)$ is the Gaussian random noise with zero mean and
given pair covariance
\begin{equation}
\langle f(x)f(x') \rangle = 2D_{0} \delta(t-t')
\delta^{(d)}({\bf x}-{\bf x'}),
\label{covar}
\end{equation}
with the positive amplitude factor $D_{0}>0$.\footnote{Strictly speaking,
a nonvanishing mean value $\langle f\rangle$ should be introduced in order
to cancel a linear in-time growth of the mean value $\langle h\rangle$.
Once we are interested in quantities like (\ref{scaling}) that involve only
differences of the fields, both the mean values can be simultaneously
ignored.}

To be precise, the model (\ref{KPZ}), (\ref{covar}) had appeared for the
first time in the seminal paper by Forster, Nelson and Stephen \cite{FNS}
in terms of the purely longitudinal (solenoidal) vector field
$u_{i}=\partial_{i} h$. Then, for $\lambda_{0}=-1$ it represents the
$d$-dimensional generalization of the Burgers equation. It can also be
mapped onto a model of directed polymers in random media and on a model
of Bose many-particle system with attraction; see e.g. \cite{Burg}.

Actually, the first two terms on the right-hand side of (\ref{KPZ}) are just
the simplest local ones that respect the symmetries $h\to h+$const and
$O(d)$. Thus the KPZ model arises ubiquitously in description of many
nonequilibrium, disordered and driven diffusive systems. Then the field
$h$ can have different meanings. For example, in \cite{Uni} the KPZ
model and its ramifications were used to study large-scale distribution
of matter in the Universe.

A few generalizations and modifications of the original KPZ model were
introduced: random noise with finite correlation time \cite{Color},
vector or matrix field $h$ \cite{Matr}, modified form of the nonlinearity
\cite{Pav} and anisotropic modifications \cite{KimKim}. In connection to
the latter, it is also worth to mention continuous anisotropic models of
self-organized criticality \cite{Tadic}.

The powerful quantitative theory of the critical state is provided by the
field theoretic renormalization group (RG); see the monographs
\cite{Zinn,Book3} and references therein. In the RG approach, possible
universality classes are associated with IR attractive fixed points of
renormalizable field theoretic models.

The RG analysis of the KPZ model, pioneered in \cite{FNS,KPZ}, eventually
(after some misunderstanding) led to the following conclusions
\cite{10}, \cite{11}. The field theoretic version of the stochastic problem
(\ref{KPZ})--(\ref{covar}) is miltiplicatively renormalizable.
The nonlinearity $(\partial h)^{2}$ in (\ref{KPZ}) is IR irrelevant
(in the sense of Wilson) for $d>2$, logarithmic (marginal) for $d=2$
and relevant for $d<2$. Thus it can be studied within the standard
perturbative RG and the expansion in $\varepsilon\equiv 2-d$.
The corresponding RG equations
possess a nontrivial fixed point with the exponents $\chi=0$, $z=2$
(the exact relation $\chi+z=2$ is dictated by Galilean symmetry).
However, the fixed point for $\varepsilon<0$ is IR repulsive, while for
for $\varepsilon>0$ it does not lie in the physical range of the model
parameters ($D_{0},\, \varkappa_{0}>0$) and thus can hardly describe the
IR asymptotic behaviour of the problem. All these results are
``perturbatively exact,'' that is, exact in all orders of the expansion
in $\varepsilon$.

One can nevertheless assume that the KPZ model possess a hypothetical
IR attractive ``strong-coupling'' fixed point, not ``visible'' within
any kind of perturbation theory.
Then, for $d=1$, the fluctuation-dissipation theorem along with the Galilean
symmetry gives the exact values $\chi=1/2$, $z=3/2$ \cite{KPZ,FNS}.
With additional (rather nontrivial) assumptions, one derives
definite exact values for the exponents in $d=2$ and $d=3$ \cite{Quanta}.
Evidence of the existence of the strong-coupling point, provided by the
so-called functional (also referred to as ``exact'' or ``nonperturbative'')
RG \cite{Canet,Canet2}, although convincing, is numerically still
not too impressive, and the situation cannot be considered satisfactory;
some other open problems are discussed e.g. in \cite{Up,Hai}.

It is well known that the behaviour of real systems near their critical
points is very sensitive to external disturbances, gravity, finite-size
effects, presence of impurities and so on; see e.g. \cite{Ivanov,Prudnikov}
for general discussion and references. What is more, some disturbances
(randomly distributed impurities in magnets and turbulent mixing of
fluid systems) can change the type of the phase transition or give
rise to new universality classes with rich and rather exotic properties.

Investigation of the effects of various kinds of deterministic or chaotic
flows (laminar shear flows, turbulent convection and so on) on the behaviour
of the critical systems (like liquid crystals or binary mixtures near their
consolution points) has shown that the flow can destroy the usual critical
behaviour: it can change to the mean-field behaviour or, under some
conditions, to a more complex behaviour described by new non-equilibrium
universality classes \cite{Satten}--\cite{AIK}.

In this paper we study the influence of the random (turbulent) motion of
the fluid, containing dissolved particles, on the IR behaviour of the
randomly growing interface, paying special attention to the effects of
compressibility. The advection by the velocity field
$\boldsymbol{v}(x)\equiv \{v_{i}(x)\}$ is introduced by the ``minimal''
replacement
\begin{equation}
\partial_{t} h \to \nabla_{t} h \equiv \partial_{t} h
+ (v_{i}\partial_{i}) h,
\label{nabla}
\end{equation}
where $\nabla_{t}$ is the Galilean covariant (Lagrangean) derivative.

We are going to acquire preliminary qualitative understanding of what can
happen if the fluid motion is taken into account. For this reason, we
neglect possible influence of the field $h(x)$ on the dynamics of the fluid
(``passive'' advection) and model the velocity field by simple Gaussian
statistics with zero mean and prescribed pair covariance with vanishing
correlation time:
\begin{eqnarray}
\langle v_{i} (t, {\bf x}) v_{j}(t',{\bf x}')\rangle =  \delta(t-t')\,
D_{ij}({\bf x}-{\bf x}'),
\nonumber \\
D_{ij}({\bf r}) = B_{0}
\int_{k>m} \frac{d{\bf k}}{(2\pi)^{d}} \, \frac{1}{k^{d+\xi}}\,
\left\{ P_{ij}({\bf k})+\alpha Q_{ij}({\bf k})\right\} \,
\exp ({\rm i} {\bf k}\cdot {\bf r}),
\label{white}
\end{eqnarray}
known as the Kazantzev--Kraichnan ensemble; see e.g. \cite{FGV}.
Here $P_{ij}({\bf k}) = \delta_{ij} - k_i k_j / k^2$ and
$Q_{ij}({\bf k})=k_i k_j/k^2$ are the transverse and the longitudinal
projectors, respectively, $k\equiv |{\bf k}|$ is the wave number,
$B_{0}>0$ is an amplitude factor and $\alpha>0$ is an arbitrary
parameter. The case $\alpha=0$ corresponds to the incompressible fluid
($\partial _i v_{i}=0$), while the limit $\alpha \to\infty$ at fixed
$\alpha B_{0}$ corresponds to the purely potential velocity field.
The exponent $0<\xi<2$ is a free parameter which can be viewed as a kind
of H\"{o}lder exponent, which measures ``roughness'' of the velocity field;
the ``Kolmogorov'' value is $\xi=4/3$, while the ``Batchelor'' limit
$\xi\to2$ corresponds to smooth velocity. The cutoff in the
integral (\ref{white}) from below at $k=m$, where $m\equiv 1/{\cal L}$ is
the reciprocal of the integral turbulence scale ${\cal L}$, provides IR
regularization. Its precise form is unimportant; the sharp cutoff is the
simplest choice for the practical calculations.

This ensemble, although it looks simple, has attracted enormous attention
in turbulence studies because of the deep insight it offers into the origin
of intermittency and anomalous multiscaling in turbulent advection and
turbulence on the whole; see the review paper \cite{FGV} and references
therein. The RG approach to that problem is reviewed in \cite{JphysA}.
In the context of our study, it is especially important that the
Kazantzev--Kraichnan ensemble allows one to easily model compressibility,
which appears rather difficult if the velocity field is described
by Navier--Stokes equations; see e.g.~\cite{ANU,Kont}.
For a compressible fluid ($\partial _i v_{i}\ne 0$), the covariant
derivative can also be introduced in an alternative way, namely,
$\nabla_{t} h \equiv \partial_{t} h + \partial_{i} (v_{i}h) $, which is
obligatory if the field $h$ has the meaning of the density of some conserved
quantity. In our case, however, $h$ is ``not conserved'' due to the nonlinear
term in (\ref{KPZ}), and in the following we will consider only the variant
(\ref{nabla}) because it preserves the symmetry $h\to h+$const of the
original KPZ problem.

The plan of the paper is the following. In section~\ref{sec:Model} we present
the field theoretic formulation of the full stochastic problem (\ref{KPZ}),
(\ref{covar}), (\ref{white}) and  diagrammatic technique.
In section~\ref{sec:Reno} we analyze ultraviolet (UV) divergences of
the model and demonstrate its multiplicative renormalizability.
Then the RG equations, as well as equations of critical scaling, can be
derived in a standard way (section~\ref{sec:RGE}). The practical
calculation of the renormalization constants and the RG functions is
discussed in the Appendix. Fixed points of the RG equations and
possible scaling regimes are studied in section~\ref{sec:FPS}. It turns out
that, in addition to Gaussian fixed point (free field theory), purely
``kinematic'' regime (the KPZ nonlinearity is irrelevant in the sense of
Wilson) and purely KPZ fixed point (turbulent transfer is irrelevant),
the RG equations possess a fully nontrivial fixed point, in which both
the nonlinearity and the mixing are important. The corresponding critical
exponents can be calculated in the form of double expansions in $\xi$ and
$\varepsilon=2-d$; they are derived in the leading one-loop order.

The regions of IR stability of the fixed points in the parameter space
$\varepsilon$, $\xi$ and $\alpha$ are found. In particular it turns out,
that for small $\alpha$ and most realistic values $d=1$ or $2$ and $\xi=4/3$
or $2$, the IR asymptotic behaviour is governed by the kinematic fixed
point with exactly known exponents. As the degree of compressibility $\alpha$
grows, the stability region of the full-scale point is getting wider and
finally absorbs the realistic values of $\varepsilon$ and $\xi$.

Derivation of the expressions like (\ref{scaling}) requires (rather simple)
analysis of the composite fields $h^{n}(x)$; this is discussed in
section~\ref{sec:Crit}.

It should be admitted, however, that all of our practical results are
derived within the framework of a standard ``perturbative''
field-theoretic RG.
The strong-coupling IR attractive RG fixed point,
if it indeed exists, definitely survives in the full-scale model, but the
issue of its IR stability lies far beyound the scope of our study.
This problem, along with some others, is discussed in sec.~\ref{sec:Conc}.

\section{Field theoretic formulation of the model} \label{sec:Model}

Let us consider first the original KPZ model without the advection.
According to the general statement \cite{MSR}
(see also the monographs \cite{Zinn,Book3}), the stochastic problem
(\ref{KPZ}), (\ref{covar}) is equivalent to the field theoretic model
of the doubled set of fields $\Phi=\{h,h'\}$ with the action functional
\begin{equation}
{\cal S}(\Phi)=\frac{1}{2}h'D_0 h'+h'\left\{-\partial_{t}h+\varkappa_0
\partial^{2} h+ \frac{1}{2}(\partial h)^{2}\right\}
\label{act1}
\end{equation}
(we have set $\lambda_0=1$).
Here and below, all needed integrations over $x = (t,{\bf x})$ and summations
over repeated tensor indices are implied, e.g.,
\begin{equation}
h'D_0 h'={D_0}\int dt\int d{\bf x} \,\, h'(t,{\bf x})\, h'(t,{\bf x}).
\end{equation}

The field theoretic formulation means that various correlation functions and
response functions of the stochastic problem (\ref{KPZ}), (\ref{covar})
can be identified with various Green functions of the field theoretic model
with the action (\ref{act1}). In other words, they are represented by
functional averages over the full set of fields $\Phi=\{ h,h'\}$ with
the weight $\exp {\cal S}(\Phi)$.

The bare propagators in the corresponding Feynman diagrammatic techniques
are determined by the free (bilinear in the fields) part of the action
(\ref{act1}). In the frequency--momentum ($\omega$--${\bf k}$) representation
they have the forms:
\begin{eqnarray}
\langle  hh' \rangle_{0} &=& \langle h'h \rangle_{0}^{*}=
\frac{1}{-{\rm i}\omega +\varkappa_0 k^{2}},
\nonumber \\
\langle  hh \rangle_{0} &=&  \frac {D_0}
{\omega^2+\varkappa_o^2 k^4}, \quad
\langle  h'h' \rangle_{0} =  0.
\label{prop1}
\end{eqnarray}
The model has only one interaction vertex $h'(\partial h)^{2}/2$.

In the diagrammatic representation, we will denote $\langle  hh \rangle_{0}$
as a straight line and $\langle  hh' \rangle_{0}$ as a straight line with
a small stroke that corresponds to the field $h'$.

Coupling with the velocity field $\boldsymbol{v}(x)\equiv \{v_{i}(x)\}$
is introduced by the substitution (\ref{nabla}) in
(\ref{KPZ}) and thus in (\ref{act1}). The full problem is then equivalent
to the field theoretic model of the three
fields $\Phi = \{ h,h', \boldsymbol{v}\} $ with the action functional
\begin{equation}
{\cal S}(\Phi)=\frac{1}{2}h'D_0 h'+h'\left\{-\nabla_{t}h+
\varkappa_0 \partial^{2} h+
\frac{1}{2}(\partial h)^{2}+f\right\}+ {\cal S}_{\boldsymbol{v}}.
\label{act2}
\end{equation}
The last term corresponds to the Gaussian averaging over the field
$\boldsymbol{v}$ with the correlator (\ref{white}):
\begin{eqnarray}
{\cal S}_{\boldsymbol{v}} = -\frac{1}{2}\int dt\int d{\bf x}\int d{\bf x'}
v_{i} (t,{\bf x}) D^{-1}_{ij}({\bf x}-{\bf x}')
v_{j}(t, {\bf x'}),
\label{Sv}
\end{eqnarray}
where $D^{-1}_{ij}$ is the inverse to the integral operation $D_{ij}$
from (\ref{white}).

Thus the Feynman diagrams for the full model (\ref{act2}) involve,
in addition to (\ref{prop1}), the new propagator (\ref{white})
and the new vertex $-h'(v\partial)h$.

The role of the coupling constants in the ordinary perturbation theory
is played by the two parameters
\begin{equation}
g_{0} = D_{0}/\varkappa_0^3 \sim {\Lambda}^{\varepsilon} , \quad
w_{0} = B_{0}/\varkappa_0\sim {\Lambda}^{\xi}.
\label{charges}
\end{equation}
The last relations follow from the dimensionality considerations
(see the next section) and define the typical UV momentum scale
$\Lambda$.

\section{UV divergences and renormalization} \label{sec:Reno}

It is well known that the analysis of UV divergences is based on the analysis
of the canonical dimensions (``power counting''); see, e.g.,
\cite{Zinn,Book3}.
The dynamic models of the type (\ref{act2}) have two independent
scales: the time scale $T$ and the length scale $L$
(as opposed to single-scale static models).

Thus the canonical dimension of some quantity $F$ (a field or a parameter in
the action functional) can be completely descibed by two numbers, the
frequency dimension $d_{F}^{\omega}$ and the momentum dimension $d_{F}^{k}$:
\[[F] \sim [T]^{-d_{F}^{\omega}} [L]^{-d_{F}^{k}}.\]
They  are found from the obvious normalization conditions
\[ d_k^k=-d_{\bf x}^k=1,\ d_k^{\omega} =d_{\bf x}^{\omega }=0,\
d_{\omega }^k=d_t^k=0,\  d_{\omega }^{\omega }=-d_t^{\omega }=1, \]
and from the requirement
that each term of the action functional be dimensionless (with
respect to the momentum and frequency dimensions separately).
Then, based on $d_{F}^{k}$ and $d_{F}^{\omega}$,
one can introduce the total canonical dimension
$d_{F}=d_{F}^{k}+2d_{F}^{\omega}$
(in the free theory, $\partial_{t}\propto\partial^{2}$). In the theory of
renormalization of dynamical models this factor plays the same role as
the conventional (momentum) dimension does in static problems;
see Chap.~5 of \cite{Book3}.

Canonical dimensions of the fields and parameters in the model (\ref{act2})
are presented in table~\ref{table1}.
It also includes renormalized parameters (the ones without the subcript
``o'') and the renormalization mass $\mu$ that will be inroduced later on.

\begin{table}
\caption{Canonical dimensions of the fields and parameters in the model
(\ref{act2}).}
\label{table1}
\begin{tabular}{c|c|c|c|c|c|c|c|c|c|c|} \hline
$F$ & $h$ & $h'$ & $\boldsymbol{v}$ & $\varkappa_{0},\varkappa $ & $D_{0}$ &  $g_{0}$ &
 $B_{0}$ & $w_{0}$ & $g,w,\alpha$ & $m,\mu,\Lambda$
\\ \hline
$d_{F}^{\omega }$ & $1$ & $-1$ & $1$ & $1$ & $3$ & $0$ & $1$ & $0$ &
$0$ & $0$
\\ \hline
$d_{F}^{k}$ & $-2$ & $d+2$ & $-1$ & $-2$ & $-d-4$ & $
2-d\equiv\varepsilon$ &
$-2+\xi$ & $\xi$ & $0$ & $1$
\\ \hline
$d_{F}$ & $0$ & $d$ & $1$ & $0$ & $\varepsilon$ &
$\varepsilon$ & $\xi$ &
$\xi$ & $0$ & $1$
\\ \hline
\end{tabular}
\end{table}

From table~\ref{table1} it follows the model is logarithmic at $d=2$
and $\xi=0$, when the both coupling constants $g_0$ and $w_0$
simultaneously become dimensionless. Hence, the UV divergences in the Green
functions manifest themselves as poles in $\varepsilon=2-d$, $\xi$ and,
in general, in all their linear combinations.

The total canonical dimension of an arbitrary 1-irreducible Green function
$\Gamma = \langle\Phi \cdots \Phi \rangle_{\rm 1-ir}$ with $\Phi=\{h,h',v\}$
in the frequency--momentum representation is given by the relation:
\begin{equation}
d_{\Gamma}=d+2-d_h N_h-d_{h'}N_{h'}-d_v N_v,
\label{dGamma}
\end{equation}
where $N_h,N_{h'},N_v$ are the numbers of corresponding fields entering
into the function $\Gamma$; see, e.g., \cite{Book3}.

The total dimension $d_{\Gamma}$ in the logarithmic theory (i.e., at
$\varepsilon=\xi=0$) is the formal index of the UV divergence:
$\delta_{\Gamma}=d_{\Gamma}|_{\varepsilon=\xi=0}$.
The superficial UV divergences, whose removal requires counterterms, can be
present only in those functions $\Gamma$ for which $\delta_{\Gamma}$ is
a non-negative integer. The counterterm is a polynomial in frequencies and
momenta of degree $\delta_{\Gamma}$ (provided the convention that
$\omega\propto k^2$ is implied).

If, for some reason, a number of external momenta occurs as an overall
factor in all diagrams of a certain Green function, the real index of
divergence $\delta_{\Gamma}'$ will be smaller than $\delta_{\Gamma}$ by
the corresponding number of unities. This is exactly what happens in our
model: the field $h$ enters the vertices $h'(\partial h)^{2}$ and
$h'(v\partial)h$ only in the form of spatial derivatives. Thus any appearance
of $h$ in some function $\Gamma$ gives such an external momentum, and the
real index of divergence is given by the expression
$\delta_{\Gamma}'= \delta_{\Gamma} - N_{h}$. Furthermore, $h$ can appear
in the corresponding counterterm only in the form of derivative.

From table~\ref{table1} and the expression (\ref{dGamma}) one obtains:
\begin{equation}
\delta_{\Gamma}'= \delta_{\Gamma} - N_{h}=4 - N_{h} - 2N_{h'} - N_{v}.
\label{IndeX}
\end{equation}

\begin{figure}[h!]
\center{\includegraphics[width=1.7cm]{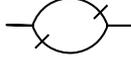}}
\caption{A diagram of the vanishing function $\langle hh\rangle_{1-ir}$
with a closed circuit of two retarded propagators.}
\label{ret1}
\end{figure}

In dynamical models, all the 1-irreducible Green functions without the
response fields vanish identically (their diagrams always involve closed
circuits of retarded lines); see, e.g., \cite{Book3}. The sample
diagram is shown in  Fig.~\ref{ret1}. Thus in (\ref{IndeX})
it is sufficient to consider the case $N_{h'}>0$.

Then straightforward analysis of the expression (\ref{IndeX}) shows that
superficial UV divergences can be present only in the following
1-irreducible functions:
\begin{eqnarray}
\langle  h'h' \rangle_{1-ir} \quad (\delta_{\Gamma}=0, \delta_{\Gamma}'=0)
\quad
{\rm with\ the\ counterterm} \quad h'h',
\nonumber \\
\langle h'hh \rangle_{1-ir} \quad (\delta_{\Gamma}=2, \delta_{\Gamma}'=0)
\quad
{\rm with\ the\ counterterm} \quad h'(\partial h)^2,
\nonumber \\
\langle h'h \rangle_{1-ir} \quad (\delta_{\Gamma}=2, \delta_{\Gamma}'=1)
\quad
{\rm with\ the\ counterterm} \quad h'\partial^2 h,
\nonumber \\
\langle h'hv \rangle_{1-ir} \quad (\delta_{\Gamma}=1,\delta_{\Gamma}'=0)
\quad
{\rm with\ the\ counterterm} \quad  h'(v\partial)h,
\nonumber \\
\langle h' \rangle_{1-ir} \quad (\delta_{\Gamma}=2, \delta_{\Gamma}'=2)
\quad
{\rm with\ the\ counterterm} \quad  h',
\nonumber \\
\langle h'v \rangle_{1-ir} \quad (\delta_{\Gamma}=1, \delta_{\Gamma}'=1)
\quad
{\rm with\ the\ counterterm} \quad  h'(\partial v),
\nonumber \\
\langle h'vv \rangle_{1-ir} \quad (\delta_{\Gamma}=0, \delta_{\Gamma}'=0)
\quad
{\rm with\ the\ counterterm} \quad  h'v^{2}.
\end{eqnarray}

Some additional considerations reduce the number of the counterterms.

The action of the KPZ model is invariant with respect to the transformation
\begin{eqnarray}
h(t,{\bf x}) \to h(t,{\bf x}+\boldsymbol{u}t) -
\boldsymbol{u}\cdot {\bf x}, \quad
h'(t,{\bf x}) \to h'(t,{\bf x}+\boldsymbol{u}t)
\label{Gali1}
\end{eqnarray}
with an arbitrary constant parameter $\boldsymbol{u}$. This invariance,
which becomes the Galilean symmetry in terms of the vector
field $\partial_{i}h$, is violated in the full model (\ref{act1}).
However, the latter possesses another kind of the Galilean symmetry,
namely,
\begin{eqnarray}
h(t,{\bf x}) \to h(t,{\bf x}+\boldsymbol{u}t), \
h'(t,{\bf x}) \to h'(t,{\bf x}+\boldsymbol{u}t) ,
\nonumber \\
\boldsymbol{v} (t,{\bf x}) \to \boldsymbol{v} (t,{\bf x}+\boldsymbol{u}t)
- \boldsymbol{u}
\label{Gali2}
\end{eqnarray}
(it is important here that our velocity field is not correlated in time).
This symmetry requires that the monomial $h'(v\partial)h$ enter the
counterterms only in the form of invariant combination
$h'\nabla_{t}h= h'\partial_{t}h+h'(v\partial)h$. The first term,
however, is forbidden by the real index (\ref{IndeX}): the field $h$
appears in it without the spatial derivative. Thus the second term is
also forbidden (the cancellation of divergent terms from different diagrams
can be checked in practical calculation, too). The Galilean symmetry
also rules out the monomial $h'v^{2}$.

\begin{figure}[h!]
\center{\includegraphics[width=1.3cm]{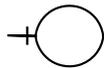}}
\caption{The one-loop ``tadpole'' diagram from the function
$\langle h' \rangle_{1-ir}$.}
\label{ret2}
\end{figure}

The counterterm $\propto h'$, stemming from the function
$\langle h' \rangle_{1-ir}$, in terms of the original stochastic problem
renormalizes the mean value of the random noise $\langle f \rangle$.
This is illustrated by the ``tadpole'' diagram in Fig.~\ref{ret2}:
it represents the one-loop contribution to the mean value
$\langle \partial h\partial h \rangle$. As already mentioned
(see the footnote on p.~2), these two mean values can simultaneously
be ignored. In this respect, the contribution from
$\langle h' \rangle_{1-ir}$ is similar to the shift of critical temperature
in models of equilibrium critical behaviour.

The counterterm of the form $h'\partial_{i}v_{i}$, stemming from the Green
function $\langle h'v \rangle_{1-ir}$, also requires special discussion.
It vanishes identically for the incompressible case, where
$\partial_{i}v_{i}=0$. However, the practical one-loop calculation shows
that it is absent in the general case ($\alpha\ne0$). One can give
some arguments that this is true in all orders of perturbation theory;
see section~\ref{sec:OL}. Since our present analysis is restricted to the
one-loop calculations, we will not take this term into account in the
following discussion.

Then we are left with the three counterterms of the form $h'h'$,
$h'\partial^{2}h$ and $h'(\partial h)^{2}$. All these terms are present
in the action (\ref{act2}) making our model multiplicatively renormalizable.
The renormalized action then can be written in the form:
\begin{eqnarray}
{\cal S}_{R}(\Phi) = \frac{1}{2}Z_1 h'D h'+h'\left\{-\partial_{t}h
+\varkappa Z_2\partial^{2} h+
\frac{1}{2}Z_3(\partial h)^{2}+f\right\}+ {\cal S}_{\boldsymbol{v}},
\label{RenAct}
\end{eqnarray}
Here $g$, $w$ and $\varkappa$ are renormalized analogs of the bare
parameters, and the functional ${\cal S}_{\boldsymbol{v}}$ from (\ref{Sv})
should also be expressed in renormalized variables.
The renormalization constants $Z_i$ depend only on the completely
dimensionless parameters $g, w, \varkappa, \alpha$ and absorb the poles
in $\varepsilon$ and $\xi$.

The renormalized action (\ref{RenAct}) is obtained from the original
one (\ref{act2}) by the renormalization of the fields $h\rightarrow Z_h h$
and $h'\rightarrow Z_{h'} h'$ and of the parametrs:
\begin{eqnarray}
\varkappa_{0} = \varkappa Z_{\varkappa}, \quad
g_{0} = g \mu^{\varepsilon} Z_{g},      \quad
w_{0} = w \mu^{\xi} Z_{w} .
\label{Multy}
\end{eqnarray}
The amplitudes $D$ and $B$ are expressed in renormalized parameters as
follows:
\begin{eqnarray}
D = g \varkappa^3\mu^{\varepsilon}, \quad
B = w \varkappa\mu^{\xi}.
\end{eqnarray}

The renormalization constants in Eqs. (\ref{RenAct}) and (\ref{Multy})
are related as follows:
\begin{eqnarray}
Z_{g} = Z_{1} Z_{2}^{-3} Z_3^2, \quad
Z_{\varkappa}= Z_{2}, \quad
Z_{h} = Z_{3}, \quad
Z_{h'} = Z_{3}^{-1}, \nonumber \\
Z_h Z_{h'}=1 , \quad Z_v=1, \quad Z_{w}Z_{\varkappa} = 1.
\label{ZZ}
\end{eqnarray}
The first two relations in the second line follow from the absense of the
counterterm $h'\nabla_t h$, the last relation follows from the absense
of renormalization of the term ${\cal S}_v$.

The renormalization constants $Z_{1}$--$Z_{3}$ are calculated directly from
the diagrams, then the constants in (\ref{Multy}) are found from
(\ref{ZZ}). The renormalization constants can be found from the
requirement that the Green functions of the renormalized model
(\ref{RenAct}) be UV finite when expressed in renormalized variables.
In our case this means that the Green functions are UV finite at
$\varepsilon,\xi\rightarrow 0$.
The calculation in the first order in $g$ and $w$ (one-loop approximation)
gives (see section~\ref{sec:OL} for the details):
\begin{eqnarray}
Z_{1} = 1 - \frac{1}{8\varepsilon}\hat g-\frac{\alpha}{2\xi}\hat w, \quad
Z_{2} = Z_{3}=1 - \frac{\hat w}{\xi} \frac{d-1+\alpha}{2d}, \quad
\label{Z}
\end{eqnarray}
where $\hat g= g S_{d}/(2\pi)^d$, $\hat w= w S_{d}/(2\pi)^d$, and
$S_{d}=2\pi^d/\Gamma(d/2)$ is the area of the unit sphere in $d$
dimensions.

If the minimal subtraction (MS) scheme is employed, the renormalization
constants must have the forms ``$Z = 1+$ only poles in $\varepsilon$ and
$\xi$'' (and in higher orders in their linear combinations). Then, strictly
speaking, in our one-loop accuracy we have to replace $d=2-\varepsilon\to2$
in the above expressions. However, for some time we will keep them in the
form (\ref{Z}): then some exact results for the special cases will be
derived; see section~\ref{sec:FPS}. A similar renormalization scheme,
where the dimension $d$ is kept in ``geometrical factors'' stemming from
contractions of various projectors, was earlier used in \cite{10}; its
validity and equivalence to the MS scheme was demonstrated in \cite{100}.

\section{RG equations and RG functions} \label{sec:RGE}

Consider briefly an elementary derivation of the RG equations; detailed
exposition can be found in the monographs \cite{Zinn,Book3}.
The RG equations are written for the renormalized Green functions
$G_{R} =\langle \Phi\cdots\Phi\rangle_{R}$. They differ from
the original (unrenormalized) Green functions $G$ by overall numerical
factors (due to rescaling of the fields) and by different choice of the
parameters ($e,\mu$ instead of $e_{0}$). Thus the renormalized Green
functions can be equally used for analyzing the critical behaviour.
The relation $S_{R} (Z_{\Phi}\Phi,e,\mu) = S(\Phi,e_{0})$ between the
functionals (\ref{act2}) and (\ref{RenAct}) yields the relations
\begin{equation}
G(e_{0},\dots) = Z_{h}^{N_{h}} Z_{h'}^{N_{h'}}
G_{R}(e,\mu,\dots).
\label{multi}
\end{equation}
between the Green functions. Here, as above, $N_{h}$ and
$N_{h'}$ are the numbers of corresponding fields
entering into $G$ (we take into account that in our model $Z_{v}=1$);
$e_{0}=\{g_{0}, \varkappa_{0}, w_{0} \}$ is a full set of
bare parameters and $e=\{ g, \varkappa, w  \}$ are their renormalized
counterparts; the ellipsis stands for the other arguments (times,
coordinates, momenta etc.).

We use $\widetilde{\cal D}_{\mu}$ to denote the differential operation
$\mu\partial_{\mu}|_{e_0}$. When expressed in the renormalized variables
it looks as folows:
\begin{equation}
{\cal D}_{RG}\equiv {\cal D}_{\mu} + \beta_{g}\partial_{g} +
\beta_{w}\partial_{w}  -
\gamma_{\varkappa}{\cal D}_{\varkappa},
\label{RG2}
\end{equation}
where ${\cal D}_{x}\equiv x\partial_{x}$ for any variable
$x$. The anomalous dimensions $\gamma$ are defined as
\begin{equation}
\gamma_{F}\equiv \widetilde {\cal D}_{\mu} \ln Z_{F} \quad
{\rm for\ any\ quantity} \ F,
\label{RGF1}
\end{equation}
and the $\beta$ functions for the two dimensionless couplings $g$ and $w$ are
\begin{equation}
\beta_{g} \equiv \widetilde {\cal D}_{\mu} g = g\,[-\varepsilon-\gamma_{g}],
\quad
\beta_{w} \equiv \widetilde {\cal D}_{\mu} w = w\,[-\xi-\gamma_{w}],
\label{betagw}
\end{equation}
where the second equalities come from the definitions and the
relations (\ref{Multy}).

In order to derive the basic RG differential equations we apply to the both
sides of the equality (\ref{multi}) the operation $\widetilde{\cal D}_{\mu}$:
\begin{equation}
\left\{ {\cal D}_{RG} + N_{h} \gamma_{h} +
N_{h'} \gamma_{h'} \right\}
\,G_{R}(e,\mu,\dots) = 0.
\label{RG1}
\end{equation}

At last, equations (\ref{ZZ}) yield the following relations between
the anomalous dimensions (\ref{RGF1}):
\begin{eqnarray}
\gamma_{h}= \gamma_{3}, \quad \gamma_{h'} = -\gamma_{3}, \quad
\gamma_{w} = -\gamma_{\varkappa}, \quad \gamma_{v} = 1,
\nonumber \\
\gamma_{\varkappa} = \gamma_{2}, \quad
\gamma_{g} = \gamma_{1}-3\gamma_{2}+2 \gamma_{3} .
\label{gadf}
\end{eqnarray}

The anomalous dimension corresponding to a given renormalization constant
$Z_{F}$ can be found from the relation
\begin{equation}
\gamma_{F} = \left(\beta_{g} \partial_{g}+\beta_{w} \partial_{w}\right)
\ln Z_{F} \simeq  - \left(\varepsilon {\cal D}_{g}+\xi
{\cal D}_{w}\right) \ln Z_{F},
\label{GfZ}
\end{equation}
obtained from the definition (\ref{RGF1}), expression (\ref{RG2}) for the
operation $\widetilde {\cal D}_{\mu}$ in renormalized variables, and the
fact that the renormalization constants depend only on the two completely
dimensionless coupling constants $g$ and $w$. In the second part of
the relation, we retained only
the leading-order terms in the $\beta$ functions (\ref{betagw}) as it is
sufficient for the first-order approximation. Utilizing the MS scheme,
in the one-loop approximation from the explicit expressions (\ref{Z}) one
finds:
\begin{eqnarray}
\gamma_{1} = \hat{g}/8+\alpha\hat{w}/2,
\quad
\gamma_{2} = \gamma_{3} = \hat{w}\frac{d-1+\alpha}{2d},
\label{gammasE}
\end{eqnarray}
with $\hat g$ and $\hat w$ defined earlier (\ref{Z}) and the corrections
of order $\hat{g}^{2}$, $\hat{w}^{2}$, $\hat w\hat g$ and higher.

\section{Fixed points and scaling regimes} \label{sec:FPS}

It is well known that the long-time large-distance asymptotic behaviour
of a renormalizable field theory is determined by IR attractive fixed
points of the corresponding RG equations. In general, the
coordinates of possible fixed points are found from the
requirement that all the $\beta$
functions vanish. In the model (\ref{act2}) the coordinates
$g_{*}$, $w_{*}$ are determined by the two equations
\begin{equation}
\beta_{g} (g_{*},w_{*}) = 0, \quad \beta_{w} (g_{*},w_{*})=0 ,
\label{points}
\end{equation}
with the $\beta$ functions given in (\ref{betagw}).
The type of a fixed point is determined by the matrix
\begin{equation}
\Omega=\{\Omega_{ij}=\partial\beta_{i}/\partial g_{j}\},
\label{OmegaDef}
\end{equation}
where $\beta_{i}$ is the full set of the $\beta$ functions and
$g_{j}= \{g,w\}$ is the full set of coupling constants. For an IR
attractive fixed point the matrix $\Omega$ should be   positive, i.e.,
the real parts of all its eigenvalues should be positive.

From the relations (\ref{gadf}) we find that $\gamma_{g}=\gamma_1-\gamma_2$,
$\gamma_{w}=-\gamma_{\varkappa}=\gamma_2$, so that
\begin{eqnarray}
\beta_{g} = g\, (-\varepsilon-\gamma_g)=g\,
(-\varepsilon-\gamma_1+\gamma_2), \nonumber \\
\beta_{w} = w\, (-\xi -\gamma_w)=w\, (-\xi -\gamma_2).
\label{betas2}
\end{eqnarray}
Substituting explicit expressions (\ref{gammasE}) we arrive at the explicit
one-loop expressions for the $\beta$ functions:
\begin{eqnarray}
\beta_{g} = g\, \left\{ - \varepsilon - \frac{\hat g}{8}
+ \hat w\, \frac{(d-1)(\alpha-1)}{2d} \right\},
\nonumber \\
\beta_{w} = w\, \left\{ - \xi + \hat w\, \frac{(d-1+\alpha)}{2d}  \right\},
\label{betas22}
\end{eqnarray}
with possible corrections of the order $w^{2}$ and so on. It is worth
noting that the one-loop result (\ref{betas22}) for $\beta_{w}$ becomes
exact at $g=0$ (see e.g. \cite{JphysA}), while the result for $\beta_{g}$
becomes exact at $w=0$ (as follows from the analysis of \cite{11}).

From Eqs. (\ref{points}) and (\ref{betas2}) one finds that there are four
different fixed points in our model. As $\partial_g \beta_w=0$, the matrix
$\Omega$ is triangular in every case and its eigenvalues are simply given
by the diagonal elements
$\Omega_{g} = \partial \beta_{g} / \partial g$ and
$\Omega_{w} = \partial \beta_{w} / \partial w$.

The fixed points are as follows:

\

\noindent 1. Gaussian (free) fixed point:
$g_{*}=w_{*}=0$;  $\Omega_{g} = -\varepsilon$,  $\Omega_{w} = -\xi$
(all these expressions are exact).

\

\noindent 2. $w_{*}=0$ (exact result to all orders),
$\hat g_{*}=-8\varepsilon$;
$\Omega_{g} = \varepsilon$,  $\Omega_{w} = -\xi$.

\

This fixed points corresponds to pure KPZ model: although the interaction
with the velocity field is present, it does not affect the leading-order
IR asymptotic behaviour (it is irrelevant in the sense of Wilson).

\

\noindent 3. $g_{*}=0$ (exact),
$\hat w_{*}= \displaystyle \frac{2d\, \xi}{(d-1+\alpha)}$;
$\Omega_{g} = -\varepsilon+\xi-\displaystyle
\frac{d\alpha\,\xi}{(d-1+\alpha)}$,
$\Omega_{w} = \xi$ (exact).

\

This fixed points corresponds to pure Kraichnan model with small-scale
stirring; the KPZ nonlinearity is irrelevant in the sense of Wilson.

\

\noindent 4. $\hat g_{*}=8 \Biggl\{-\varepsilon+\xi- \displaystyle
\frac{d\alpha\,\xi}{(d-1+\alpha)} \Biggr\}$,
$\hat w_{*}=\displaystyle \frac{2d\xi}{(d-1+\alpha)}$;
$\Omega_{g} = \varepsilon-\xi+\displaystyle \frac{d\alpha\,\xi}
{(d-1+\alpha)}$, $\Omega_{w} = \xi$ (exact).

\

This fixed point corresponds to a new nontrivial IR scaling regime
(universality class), in which the nonlinearity of the model
(\ref{act2}) and the turbulent mixing are simultaneously important.

\begin{figure}[h!]
\begin{center}
\includegraphics[width=0.5\linewidth]{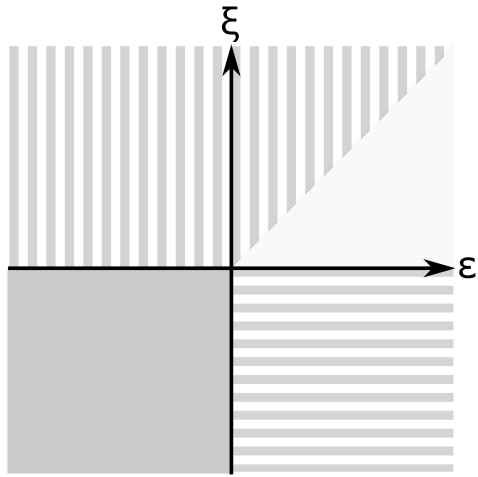}
\caption{\label{fig:pattern}
Regions of stability of the fixed points in the model (\protect\ref{act2}).}
\end{center}
\end{figure}

In figure~\ref{fig:pattern} the regions of IR stability for all the
fixed points in the $\varepsilon$--$\xi$ plane are shown:
dark space for the Gaussian point, horizontal shading for the KPZ point,
vertical shading for the Kraichnan point and
white space for the new regime.
These regions are areas in which the eigenvalues of the matrix
(\ref{OmegaDef}) for the given fixed point are both positive.

In the one-loop approximation (\ref{betas2}), all the boundaries of the
regions of stability are straight rays. Different regions have neither gaps
nor overlaps between them. Such a pattern is typical feature of the
first-order approximations.
The boundaries $\varepsilon<0$, $\xi<0$ for point 1, $\varepsilon>0$ for
point 2 and $\xi>0$ for point 3 are exact, while the others can be affected
by the higher-order corrections, i.e., the boundaries may become curved and
gaps or overlaps may appear between the different regions of IR stability.

\begin{figure}[h!]
\begin{center}
\includegraphics[width=0.5\linewidth]{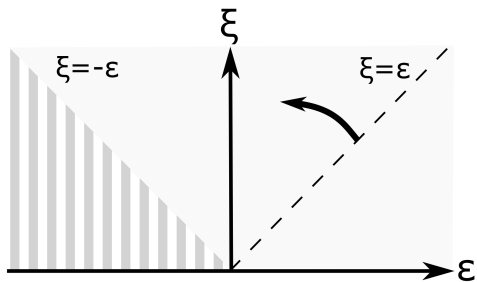}
\caption{\label{fig:pattern2}
The boundary between the regions 3 and 4 moves counterclockwise with the
growth of the compressibility $\alpha$ until it reaches the ray
$\varepsilon=-\xi$ (for $d=2$).}
\end{center}
\end{figure}

The main qualitative conclusion that can be drawn from this pattern, is
that for small $\alpha$ and most realistic values $d=1$ or $2$ and $\xi=4/3$
or $2$, the IR asymptotic behaviour is governed by the Kraichnan fixed
point. However, as the degree of compressibility $\alpha$ increases, the
stability region of the new point is getting wider and finally absorbs
the realistic values of $\varepsilon$ and $\xi$.
Indeed, the boundary between the regions 3 and 4 depend on $\alpha$.
When $\alpha$ grows, it rotates counterclockwise and, for
$\alpha\to\infty$, approaches the ray $\varepsilon=(1-d)\xi$.
Thus for $d=1$ it tends to the vertical ray $\varepsilon=0$, $\xi>0$
and for $d=2$ it tends to $\varepsilon=-\xi$, $\xi>0$; see
figure~\ref{fig:pattern2}
(note, that for $d=2$ the boundary becomes vertical at $\alpha=1$).

We recall, however, that the results concerning the boundary between the
regions
3 and 4 may be affected by the higher-order contributions. We also note
that interpretation of the fixed point 4 leads to the same difficulty as
that of the KPZ point: when it is IR attractive, the coordinate $g_{*}$
lies in the unphysical region $g_{*}<0$. We will return to this issue
in section~\ref{sec:Conc}.

\section{Critical dimensions. Critical scaling of the structure functions.
Composite fields $h^{n}(x)$.}
\label{sec:Crit}

Existence of an IR attractive fixed point implies existence of scaling
(self-similar) behaviour of the Green functions in the IR range. In this
critical scaling all the ``IR irrelevant'' parameters ($\mu$, $g$ and $w$
in our case) are kept fixed, while the ``IR relevant'' parameters
(coordinates/momenta, times/frequencies, the fields) are dilated. The
critical dimension $\Delta_{F}$ of a certain IR relevant quantity $F$ is
given by the relations (with the normalization condition $\Delta_{k} = 1$)
\begin{eqnarray}
\Delta_{F} = d^{k}_{F}+ \Delta_{\omega} d^{\omega}_{F} + \gamma_{F}^{*},
\label{dim}
\end{eqnarray}
where
\begin{eqnarray}
\Delta_{\omega} = 2 -\gamma_{\varkappa}^{*}
\label{dim2}
\end{eqnarray}
is the critical dimension of frequency, $d^{k,\omega}_{F}$ are the canonical
dimensions of $F$, given in table~\ref{table1}, and $\gamma_{F}^{*}$ is the
value of the anomalous dimension (\ref{RGF1}) at the fixed point in question:
$\gamma_{F}^{*} = \gamma_{F} (g_{*},w_{*})$; see e.g. \cite{Book3} for
detailed explanation.

From table~\ref{table1} for the dimensions of the fields we obtain
\[ \Delta_{h} = -2 + \Delta_{\omega} + \gamma_{h}^{*}, \quad
\Delta_{h'} = (d+2) - \Delta_{\omega} + \gamma_{h}^{*}. \]
Then the relations (\ref{gadf})  give
\[ \Delta_{\omega} = 2-\gamma_{2}^{*}, \quad
\Delta_{h} = -\gamma_{2}^{*} + \gamma_{3}^{*} , \quad
\Delta_{h'} = d + \gamma_{2}^{*} -\gamma_{3}^{*} . \]
Finally, substituting explicit one-loop expressions (\ref{gammasE}) yields
\begin{eqnarray}
\Delta_{h} = 0, \quad \Delta_{h'} = d, \quad \Delta_{\omega} = 2
\label{dimD}
\end{eqnarray}
for the fixed points 1 and 2 and
\begin{eqnarray}
\Delta_{h} = 0, \quad \Delta_{h'} = d, \quad \Delta_{\omega} = 2-\xi
\label{dimD2}
\end{eqnarray}
for the points 3  and 4. For the points 1--3 these expressions are exact;
see the remark below eq.~(\ref{betas22}). For the point 4 the dimensions
of the fields can be affected by the higher-order contributions with
the exact condition that $\Delta_{h} +\Delta_{h'} = d$, and the expression
for $\Delta_{\omega}$ is exact due to the second relation in (\ref{betas2}).

In the traditional notation (\ref{scaling}) one has $\Delta_{h}= - \chi$
and $\Delta_{\omega}=z$. However, the quantity $S_{n}$ in (\ref{scaling})
is not an ordinary $n$-th order Green function of the basic fields $h(x)$:
it is a sum of pair correlation functions $\langle h^{s}(x)h^{q}(0)\rangle$
of the composite fields (``composite operators'' in the quantum-field
terminology) $h^{n}(x)$. Renormalization of such quantities requires
further analysis which, however, is rather simple in the present case.

Total canonical dimension of the 1-irreducible Green function
$\Gamma =\langle F\, \Phi \dots \Phi \rangle_{1-ir}$ with one composite
operator $F$ and arbitrary number of basic fields $\Phi=\{h,h',v\}$
in our model is $d_{\Gamma}=d_{F}-d_h N_h-d_{h'}N_{h'}-d_v N_v$,
where $N_h,N_{h'},N_v $ are the numbers of corresponding fields entering
into the function $\Gamma$, $d_{h,h',v}$ are their canonical dimensions
and $d_{F}$ is the canonical dimension of $F$; see, e.g., \cite{Book3}.
The formal index of divergence is
$\delta_{\Gamma}=d_{\Gamma}|_{\varepsilon=\xi=0}$; superficial
divergences can be present in $\Gamma$ if $\delta_{\Gamma}$ is
a non-negative integer.

From table~\ref{table1} for $F=h^{n}$  we obtain $d_{F}=0$ and
$\delta_{\Gamma} = - 2 N_{h'} - N_{v}$. Thus the divergences, at first
sight, can be present in all functions
$\Gamma =\langle h^{n}\, h \dots h \rangle_{1-ir}$ with $N_{h'} = N_{v} =0$,
arbitrary $N_{h}$ and $\delta_{\Gamma} =0$. However, any nontrivial
diagram of such a function involves at least one external vertex
$h'(\partial h)^{2}$ or $h'(v\partial)h$, where the field $h$ stands under
a derivative. Thus at least external momentum appears in the diagram as an
overall factor, the real index $\delta_{\Gamma}'$ is negative and the
superficial divergence is in fact absent.

This means that all the operators $F=h^{n}$ are not renormalized and their
critical dimensions are simply given by $\Delta_{F} = n \Delta_{h}$. This
justifies the relation (\ref{scaling}) with the dimensions (\ref{dimD}),
(\ref{dimD2}).

\section{Discussion and conclusion} \label{sec:Conc}

We studied effects of turbulent mixing in the problem of randomly growing
interface. The growth was modelled by the well-known Kardar--Parisi--Zhang
stochastic equation (\ref{KPZ}), (\ref{covar}).
The turbulent velocity field was modelled by the Kraichnan's rapid-change
ensemble (\ref{white}).

The full problem can be refornulated as the
multiplicatively renormalizable model with the action functional
(\ref{act2}). Using the field theoretic RG we
show that, depending on the relation between the exponent $\xi$ and the
spatial dimension $d$, the system exhibit different types of IR behaviour,
associated with four possible fixed points
of the RG equations. In addition to known regimes
(ordinary diffusion, ordinary growth process, and passively advected scalar
field), existence of a new nonequilibrium universality class is established.

Practical calculations of the fixed point coordinates, their regions of
stability and critical dimensions were performed in the first order of the
double expansion in $\xi$ and $\varepsilon=2-d$ (one-loop approximation).

It was shown that for the incompressible fluid the most realistic values of
$\xi$ and $d$ correspond to the universality class of passive scalar field,
when the nonlinearity of the KPZ model is irrelevant.
If the degree of compressibility $\alpha$ becomes large enough,
the crossover in the IR behaviour occurs, and these values of
$d$ and $\xi$ fall into the region of stability of the new regime.

However, a few problems remain open.

For the new universality class (as well as for the KPZ model) the coordinates
of the fixed point lie in the unphysical region $g_{*}<0$, which corresponds
to the ``wrong'' negative sign of the amplitude in the pair correlator
(\ref{covar}). Thus it requires a careful physical interpretation.
In this connection one can recall, that in the Doi--Peliti
formalism \cite{Doi,Tauber}, where the original microscopic problem
is formulated in terms of the creation-annihilation operators, the terms
quadratic in the response fields can appear in the action functionals
with the negative signs; see e.g. \cite{Tauber}.

Another question is the fate of the strong-coupling fixed point of the
pure KPZ model \cite{Canet}. If it indeed exists, it definitely survives
in our problem (with the second coordinate $w_{*}=0$), but can become
unstable. On the other hand, new nonperturbative fixed points with
$w_{*}\ne0$ can appear.

In our analysis we employed the simplest Kraichnan's ensemble for the
advecting velocity field. It would be interesting to consider more
realistic models with non-Gaussianity, finite correlation time, anisotropy,
etc. This work is in progress.

\section*{Acknowledgments}

The authors are indebted to L.Ts. Adzhemyan for helpful discussion.
The authors acknowledge Saint Petersburg State University for
the research grant 11.38.185.2014.

\appendix
\section{Calculation of the renormalization constants} \label{sec:OL}

In this Appendix we shortly present  derivation of the first-order
results (\ref{Z}) for the renormalization constants. Although the one-loop
calculation is rather simple and can be accomplished in a few different
ways, it is worth to discuss it for completeness and in order to mention
some interesting subtleties specific of the model (\ref{act2}).

The renormalization constants are determined by the requirement that the
Green functions of the renormalized model (\ref{RenAct}), expressed in
renormalized variables, be UV finite (in our case, be finite at
$\varepsilon\to0$, $\xi\to0$). The full set of
constants $Z_{1}$--$Z_{3}$ can be found from the three 1-irreducible
functions: $\langle h^{'} h^{} \rangle_{\rm 1-ir}$, $\langle h^{'}
h^{'} \rangle_{\rm 1-ir}$ and $\langle h^{'}hh  \rangle_{\rm
1-ir}$. In the renormalized model, the corresponding one-loop
approximations have the forms
\begin{eqnarray}
\langle h^{'} h^{} \rangle_{\rm 1-ir} = {\rm i}\eta - \varkappa p^2
Z_{2} \ + \
\raisebox{-0.32cm}{\includegraphics[width=1.7cm]{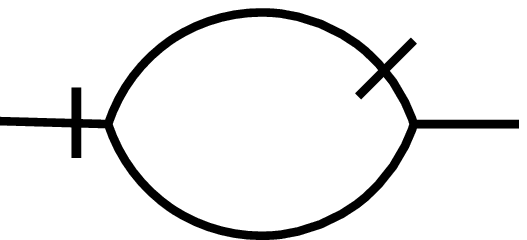}} \
+ \
\raisebox{-0.32cm}{\includegraphics[width=1.7cm]{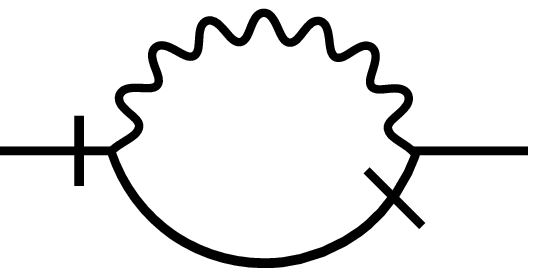}}
\ ,
\nonumber \\ {}
\label{Diagr0}
\end{eqnarray}
\begin{eqnarray}
\langle h^{'} h^{'} \rangle_{\rm 1-ir} = D Z_{1} + \frac{1}{2} \
\raisebox{-0.28cm}{\includegraphics[width=1.7cm]{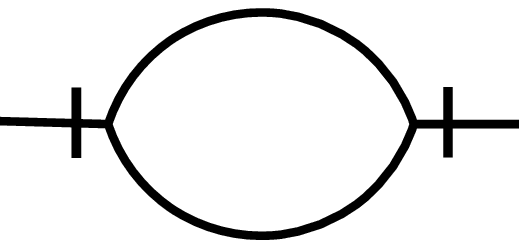}} \
+ \
\raisebox{-0.28cm}{\includegraphics[width=1.7cm]{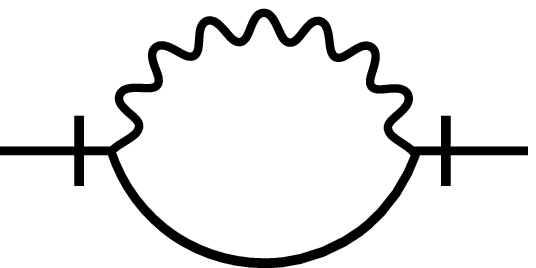}}
\ \nonumber \\ {} \label{Diagr1}
\end{eqnarray}
 and
\begin{eqnarray}
\langle h'hh \rangle_{\rm 1-ir} = \
\raisebox{-0.20cm}{\includegraphics[width=0.3cm]{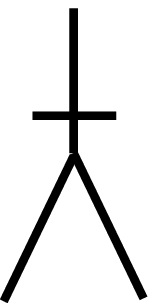}} \
+ \
\raisebox{-0.41cm}{\includegraphics[width=1.3cm]{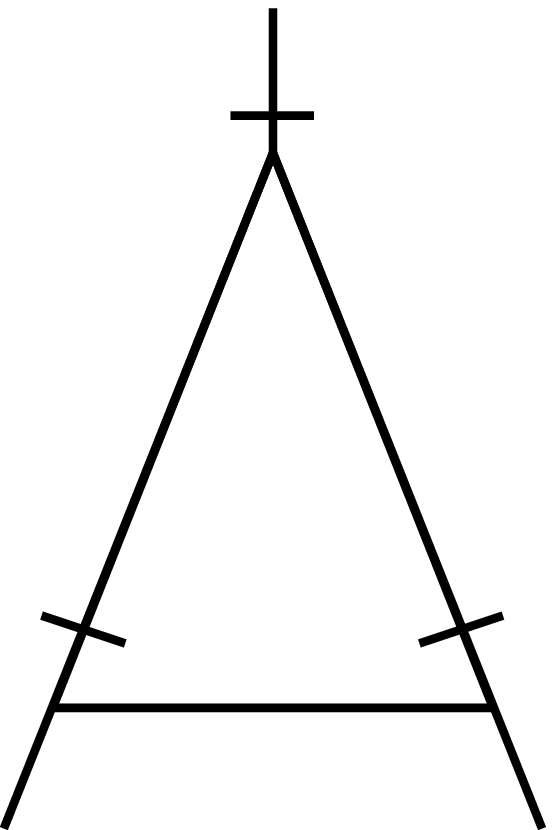}} \
+ \
\raisebox{-0.41cm}{\includegraphics[width=1.3cm]{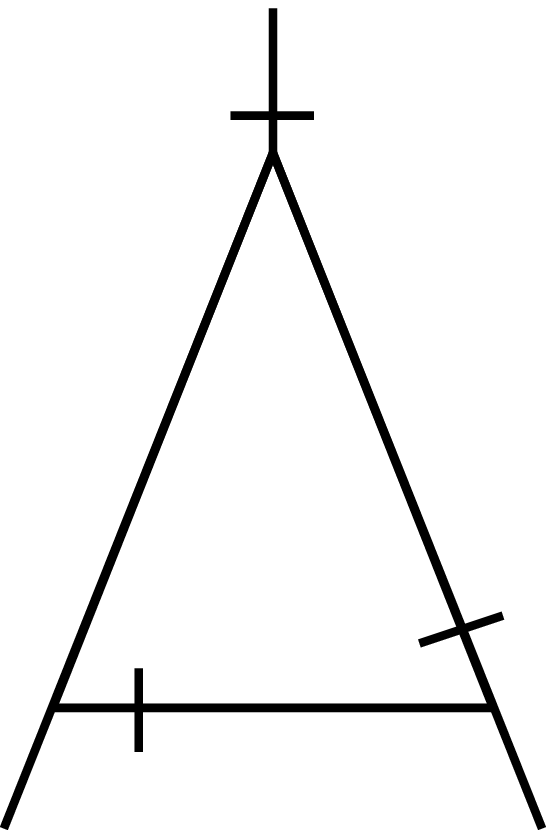}} \
+ \
\raisebox{-0.41cm}{\includegraphics[width=1.3cm]{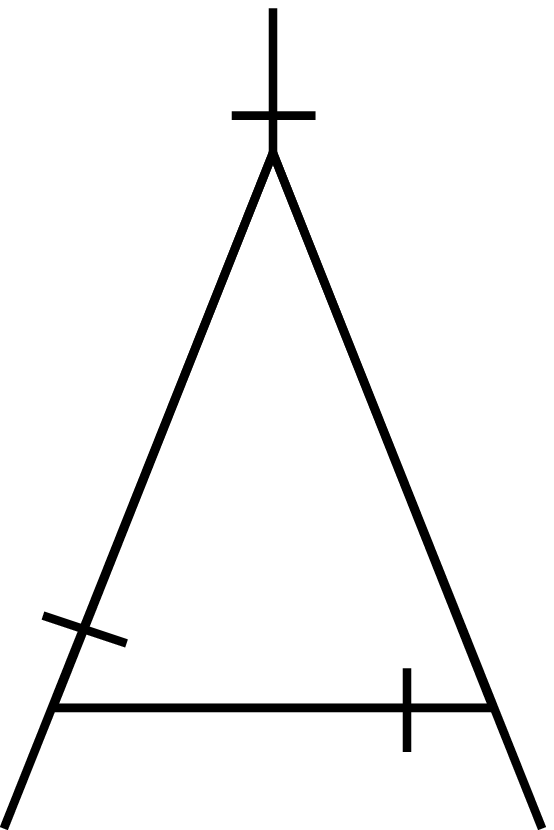}} \ +
\nonumber \\
+ \
\raisebox{-0.41cm}{\includegraphics[width=1.3cm]{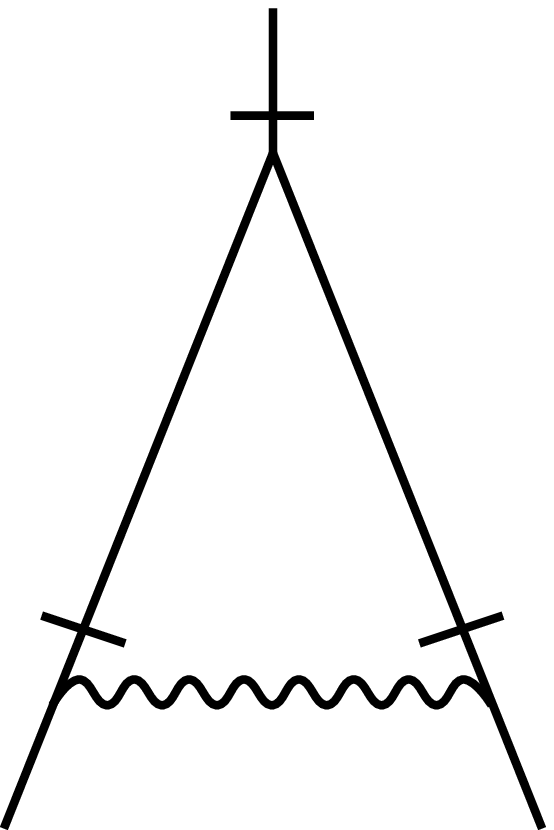}} \
+ \
\raisebox{-0.41cm}{\includegraphics[width=1.3cm]{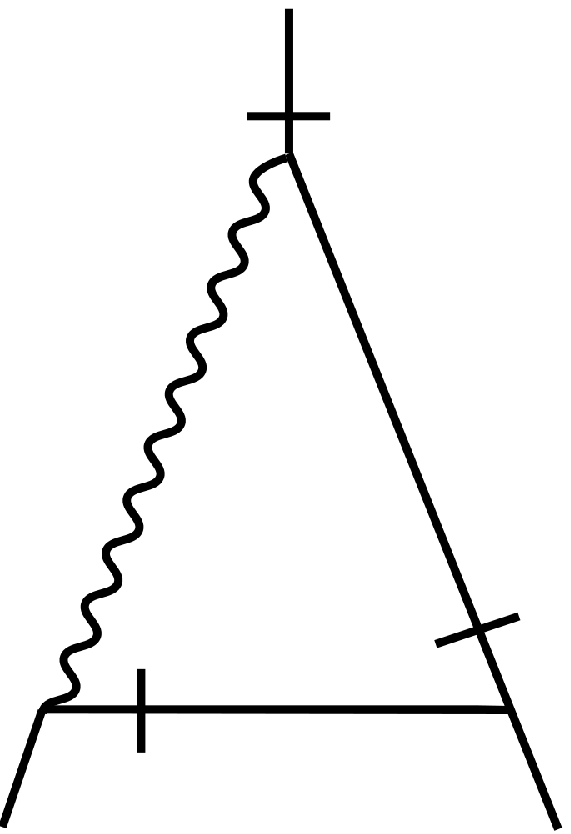}} \
+ \
\raisebox{-0.41cm}{\includegraphics[width=1.3cm]
{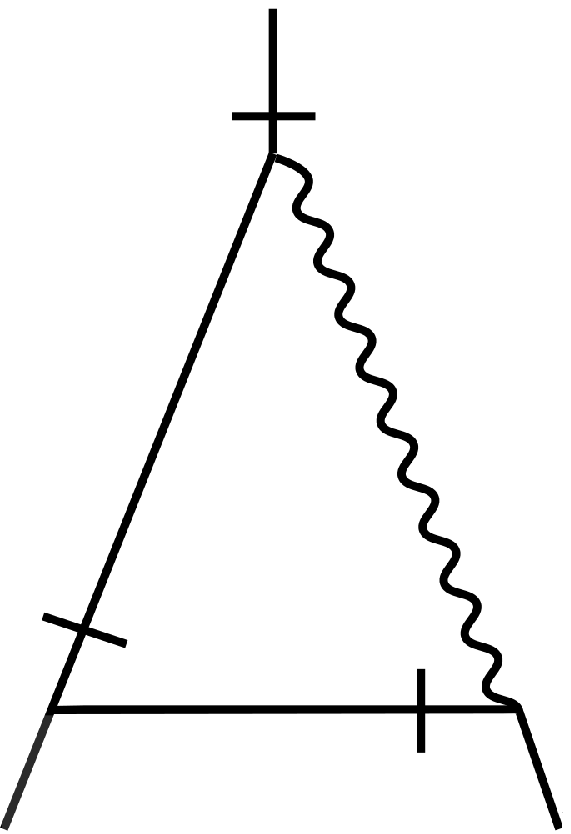}}.
\nonumber \\
\label{Diagr2}
\end{eqnarray}

Here we denoted the bare propagator $\langle  hh \rangle_{0}$ as a straight
line,  $\langle  hh' \rangle_{0}$ as a straight line with a small stroke that
corresponds to the field $h'$, and the velocity propagator as the wavy line.
In the following, we set the external frequency $\eta$ equal to 0, because
the divergent part of the function $\langle h^{'} h^{} \rangle_{\rm 1-ir}$
is proportional to $p^{2}$, where $p$ is the external momentum.

All the diagrammatic elements should be expressed in renormalized variables
using the relations (\ref{RenAct})--(\ref{ZZ}). In the one-loop
approximation, the constants $Z_i$ in the bare terms of
(\ref{Diagr0}), (\ref{Diagr1}) and (\ref{Diagr2}) should be taken
in the first order in $g$ and $w$, while in the one-loop contributions
they should simply be replaced with unities, $Z_{i}\to1$. Thus the passage
to renormalized variables in the one-loop diagrams is achieved by the
simple substitutions $\varkappa_{0} \to \varkappa$,
$g_{0} \to g\mu^{\varepsilon}$ and $w_{0} \to w\mu^{\xi}$.

The IR regularization in the diagrams involving the velocity propagator
is provided by the cutoff in the integral (\ref{white}) from below at $k=m$.
In other diagrams, IR regularization is provided by external momenta and
frequencies. We are interested, however, only in the UV divergent parts
of these diagrams (poles in $\varepsilon$ and $\xi$). Thus we will use
the following trick, which simplifies the calculation: integrations
over the momenta in all diagrams will be cut off from below at $k=m$. Then
in the logarithmically divergent functions (\ref{Diagr1}) and (\ref{Diagr2})
external momenta and frequencies can be set equal to zero, when in the
quadratically divergent function (\ref{Diagr0}) we will keep only the
$p^{2}$ term of the expansion in the external momentum  $p$.

Now the integrations over the frequency are easily performed by residues.
The resulting integrals over the momentum with the aid of the formulas
\begin{eqnarray}
\int\! {d{\bf k}} k_{i} f(k) =0, \quad
\int\! {d{\bf k}} \frac{k_{i}k_{s}}{k^{2}} f(k) =
\frac{\delta_{is}}{d}\, \int {d{\bf k}}\, f(k), \
\label{tenz}
\end{eqnarray}
where $f(k)$ is any function depending only on $k=|{\bf k}|$,
are reduced to the scalar integral
\begin{eqnarray}
J(m)= \int_{k>m} {d{\bf k}}\, \frac{1}{k^{d+y}} = S_{d}
\frac{m^{-y}}{y}
\label{ska}
\end{eqnarray}
with $S_{d}$ from (\ref{Z}). Here either $y=\varepsilon$ or $y=\xi$.

The last two diagrams in (\ref{Diagr2}) in fact vanish and thus give no
contribution to the renormalization constant. Indeed, they effectively
involve closed circuits of retarded propagators; it is crucial here that
the velocity correlator contains the $\delta$ function in time.

Direct calculation shows that the  first three diagrams in (\ref{Diagr2})
also give no contribution to $Z_{3}$ because their divergent parts
cancel each other. This is a consequence of the Galilean symmetry
(\ref{Gali1}) of the original KPZ model, which forbids the counterterm
$h'(\partial h)^{2}$ in all orders of perturbation theory.

The analytic expression for the only remaining diagram in (\ref{Diagr2})
has the form:
\begin{eqnarray}
p_{i}p_{j} \int \frac{d\omega}{(2\pi)}
\int_{k>m} \frac{d{\bf k}}{(2\pi)^{d}} \,
\frac{k^{2}}{\omega^2+\varkappa^2 k^4} \,
\frac{w\varkappa\mu^{\xi}}{k^{d+\xi}}\,
\left\{P_{ij}({\bf k})+\alpha Q_{ij}({\bf k})\right\}.
\label{D2}
\end{eqnarray}
Here the prefactor comes from the two lower vertices, the first cofactor
in the integrand comes from the upper vertex (numerator) and from the
propagators $\langle h'h\rangle_{0}$ (denominator), and the remaining
factor is the velocity correlator. Proceeding as explained above, we
finally obtain for (\ref{Diagr2}):
\begin{eqnarray}
\langle h'hh \rangle_{\rm 1-ir} = p^2 \left\{ Z_3+
\frac{\hat{w}}{\xi}\, \left(\frac{\mu}{m}\right)^{\xi}\,
\frac{(d-1+\alpha)}{2d} \, \right\}.
\label{Y}
\end{eqnarray}
The factor $(\mu/m)^{\xi}$ is UV finite: it tends to unity for $\xi\to 0$.
We can see that, in order to cancel the pole in $\xi$ in (\ref{Y}), the
renormalization constant $Z_{3}$ can indeed be chosen in the form (\ref{Z}).

Now let us turn to the renormalization constant $Z_{2}$.

The first one of the two diagrams in (\ref{Diagr0}) appears UV finite and
does not contribute to $Z_{2}$. To see this, consider the corresponding
analytic expression (up to insufficient amplitude factors):
\begin{eqnarray}
\int \frac{d\omega}{(2\pi)} \int_{k>m} \frac{d{\bf k}}{(2\pi)^{d}}
\, \frac{k_i(p+k)_i k_j p_j}{\omega^2+\varkappa^2 k^4}\
\frac{1}{-i\omega+\varkappa |{\bf k}+{\bf p}|^2} \propto
\nonumber \\
\propto \int_{k>m}\frac{d{\bf k}}{(2\pi)^{d}} \, \frac{k_i(p+k)_i k_j
p_j}{k^2(k^2+ |{\bf k}+{\bf p}|^2)}.
\label{D31}
\end{eqnarray}
We are interested in the $p^{2}$ term of its expansion in $p$. It is a sum
of two identical integrals with opposite signs. Indeed, the first one comes
from the contribution $p_{i}p_{j}k_{i}k_{j}$ in the numerator; then in the
denominator we can put ${\bf p}=0$. The resulting integrand becomes equal to
$p_i p_j k_i k_j/2k^4$. The second one comes from the contribution
$k_{i}k_{i}k_{j}p_{j}$. Then one has to expand the denominator up to the
order ${\cal O}({\bf p})$:
\begin{eqnarray}
\int_{k>m}\frac{d{\bf k}}{(2\pi)^{d}} \, \frac{ k_j p_j}{k^2+
|{\bf k}+{\bf p}|^2}\
\simeq \int_{k>m}\frac{d{\bf k}}{(2\pi)^{d}} \,
\frac{ k_j p_j}{2k^2} \left\{ 1-\frac{(\bf{p}\bf{k})}{k^2} \right\}.
\label{expanz}
\end{eqnarray}
The first term in (\ref{expanz}) vanishes because the integrand is odd
in ${\bf k}$, and the second equals to the aforementioned one up to the
minus sign.

The analytic expression for the second diagram in (\ref{Diagr0})
is as follows:
\begin{eqnarray}
\int \frac{d\omega}{(2\pi)} \int_{k>m} \frac{d{\bf k}}{(2\pi)^{d}}
\, \frac{ip_i i(p+k)_j}{k^{d+\xi}}\,
w\varkappa\mu^{\xi}\,\frac{\left\{ P_{ij}({\bf k})+\alpha Q_{ij}({\bf k})\right\}}{-i\omega+\varkappa|{\bf k}+{\bf
p}|^2}.
\label{D3}
\end{eqnarray}
Integration over $\omega$ involves the indeterminacy
\begin{eqnarray}
\int \frac{d\omega}{(2\pi)} \, \frac{1}{-{\rm i} \omega +
\varkappa|{\bf k}+{\bf p}|^2} =  \theta(0), \label{inde}
\end{eqnarray}
where $\theta(0)$ is the Heaviside step function at the origin. This
reflects the details of the velocity statistics lost in the
white-noise limit; see the discussion in \cite{FGV}. In the case
at hand, the $\delta$ function in (\ref{white}) should be
understood as the limit of a narrow function which is necessarily
{\it symmetric} in $t$, $t'$, because one deals with a pair
correlator. Thus the indeterminacy in (\ref{inde}) must be
unambiguously resolved as half the sum of the limits:
$\theta(0)=1/2$. The integrand of the resulting integral over ${\bf k}$ in
(\ref{D3}) has an odd contribution that can be dropped and the even
contribution that yields
\begin{eqnarray}
\langle h'h \rangle_{\rm 1-ir} = -\varkappa p^2
\left\{ Z_2+ \frac{\hat{w}}{\xi}\, \left(\frac{\mu}{m}\right)^{\xi}\,
\frac{(d-1+\alpha)}{2d} \,
\right\}.
\end{eqnarray}
We can see that $Z_{2}$ can be chosen in the form (\ref{Z}); in the one-loop
approximation $Z_{2}=Z_{3}$.

We will not discuss the calculation of $Z_1$ in detail: the first diagram
reduces to the scalar integral (\ref{ska}) with $y=\varepsilon$
immediately after integration
over the frequency. The analytic expression for the second diagram is
similar to (\ref{D2}) with the replacement $p_{i}p_{j}\to k_{i}k_{j}$,
so that only the contribution from the longitudinal projector survives.
This leads to the replacement  $(d-1+\alpha)/2d \to\alpha$.
Bringing all contributions together and taking into account the symmetry
coefficient $1/2$ for the first diagram, one obtains:
\begin{eqnarray}
\langle h'h' \rangle_{\rm 1-ir} = D \left\{ Z_1+ \frac{1}{8} \,
\frac{\hat{g}}{\varepsilon}
\left(\frac{\mu}{m}\right)^{\varepsilon}\,  +\alpha
\frac{\hat{w}}{2\xi}\left(\frac{\mu}{m}\right)^{\xi} \right\} .
\label{mumu}
\end{eqnarray}
We can see that the constant $Z_{1}$ that removes the poles from expression
(\ref{mumu}) can be taken in the form (\ref{Z}).

\begin{figure}[h!]
\center{\includegraphics[width=1.7cm]{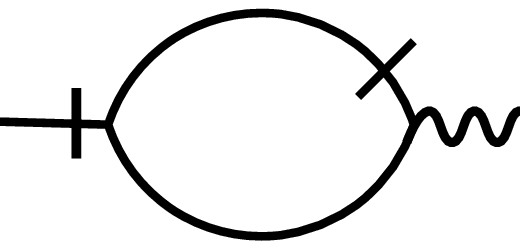}}
\caption{The one-loop contribution to the function
$\langle h' v \rangle_{1-ir}$.}
\label{vc}
\end{figure}

It remains to discuss possible UV divergence in the function
$\langle h' v \rangle_{1-ir}$ with the counterterm $h'\partial_{i}v_{i}$.
The only one-loop diagram for this function is shown in figure~\ref{vc}.
It is not difficult to see that the corresponding analytical expression
is nearly identical to that of the first diagram in the function
(\ref{Diagr0}): the latter only has additional extra factor $p_{j}$.
Thus the diagram in question is also UV finite and does not give rise to
the corresponding counterterm.

\end{document}